\documentclass[a4paper,11pt]{amsart}
\begin{document}
\hyphenation{gra-vi-ta-tio-nal re-la-ti-vi-ty Gaus-sian
re-fe-ren-ce re-la-ti-ve gra-vi-ta-tion Schwarz-schild
ac-cor-dingly gra-vi-ta-tio-nal-ly re-la-ti-vi-stic pro-du-cing
de-ri-va-ti-ve ge-ne-ral ex-pli-citly des-cri-bed ma-the-ma-ti-cal
de-si-gnan-do-si coe-ren-za pro-blem gra-vi-ta-ting geo-de-sic
per-ga-mon cos-mo-lo-gi-cal gra-vity cor-res-pon-ding
de-fi-ni-tion phy-si-ka-li-schen ma-the-ma-ti-sches ge-ra-de
Sze-keres con-si-de-red theo-ry ap-proxi-ma-te}
\title[On Gravitational Motions]
{{\bf On Gravitational Motions}}

\author[Angelo Loinger]{Angelo Loinger}
\address{A.L. -- Dipartimento di Fisica, Universit\`a di Milano, Via
Celoria, 16 - 20133 Milano (Italy)}
%\date{}
\email{angelo.loinger@mi.infn.it}
%\thanks{To be published on \emph{Spacetime \& Substance.}}

\vskip0.50cm

\begin{abstract}
\textbf{1}. Introduction. -- \textbf{2, 2bis}. Exact GR: a new
proof of the geodesic character of all motions of bodies that
interact only gravitationally. -- \textbf{3 to 7}. Linearized
approximation of GR: a detailed illustration of its real meaning
and of its shortcomings. -- \textbf{8}. Further considerations. --
\textbf{9}. A significant isomorphism. -- Historical appendix.
\end{abstract}

\maketitle

%% A new proof of the geodesic character of all motions of bodies
%% that interact only gravitationally - and a detailed illustration
%% of the real meaning of the linearized approximation of general relativity.

%%    \ddot{\Re} + \f\textbf{5}
%% {\kappa}{6}\Re \rho=0 , \tag{7'}
%% ``mechanisms'' \textrm{d} \`a
%% \cite{1}
%% eqs.(\ref{eq:six})


\vskip0.60cm \footnotesize
\begin{flushright} In memoria di Tullio Levi-Civita
(1873-1941).
\end{flushright}

\normalsize

\vskip0.60cm \noindent \textbf{1.} -- In a continuous and
incoherent ``cloud of dust'', the elements of which interact
\emph{only} gravitationally, the motions of these elements are
\emph{geodesic}, as it follows from Einstein field equations (see,
\emph{e.g.}, \cite{1}). Consequently, \emph{no} GW is emitted.

\par Now, this result can be obtained \emph{quite generally},
\emph{i.e.} without specifying the nature of matter tensor
$T_{jk}$, $(j,k=0,1,2,3)$, as I shall now prove (sects.
\textbf{2}, \textbf{2bis}): all gravitational motions are
geodesic.

\par The sects. \textbf{3} to \textbf{7} regard an analysis of the
\emph{linearized} approximation of the exact GR. A preliminary and
suitable advice: it does \emph{not} concern the instances of
motions of test-particles and light-rays in ``external'',
``rigidly given'', and weak gravitational fields. For a conceptual
dispatch of these cases, the consideration at the end of sect.
\textbf{3} is clearly sufficient; it embraces also the
computations of Thirring-Lense effects.

\vskip1.20cm \noindent \textbf{2.} -- Let $\textrm{d}s^{2}=
g_{jk}(x) \, \textrm{d}x^{j}\textrm{d}x^{k}$ be the spacetime
interval of the pseudo-Riemaniann manifold pertinent to a system
of bodies that interact only gravitationally. Denoting with
$q^{j}(\tau)$, $(j=0,1,2,3)$, the translational coordinates of one
of them as functions of proper time $\tau$, we have that

\begin{equation} \label{eq:one}
\mathcal{L} := g_{jk} \, [q(\tau)] \,
\frac{\textrm{d}q^{j}}{\textrm{d}\tau} \,
\frac{\textrm{d}q^{k}}{\textrm{d}\tau} = c^{2}
\end{equation}

is a \emph{\textbf{first integral}} of Lagrange equations

\begin{equation} \label{eq:two}
\frac{\partial \mathcal{L}}{\partial q^{j}} -
\frac{\textrm{d}}{\textrm{d}\tau} \, \frac{\partial
\mathcal{L}}{\partial \, (\textrm{d}q^{j} /  \textrm{d} \tau)} = 0
\quad.
\end{equation}

\par Now -- as it is well known -- eqs. (\ref{eq:two}) coincide with
\emph{geodesic} equations

\begin{equation} \label{eq:three}
\frac{\textrm{d}^{2}q^{j}}{\textrm{d}\tau^{2} } + \Gamma^{j}_{mn}
\, \frac{\textrm{d}q^{m}}{\textrm{d}\tau } \, \,
\frac{\textrm{d}q^{n}}{\textrm{d}\tau }
  = 0
\quad.
\end{equation}

\par In the last analysis, this means that the geodesic character
of the motions of our gravitating bodies is implicitly contained
in the pseudo-Riemannian structure of $\textrm{d}s^{2}$, from
which it follows the existence of the first integral
$\mathcal{L}=c^{2}$. This result is \emph{independent} of the
precise expressions of the components $g_{jk}$'s of the metric
tensor, that are determined by Einstein equations (plus the
boundary conditions).

\par It is certain that in a geodesic motion \emph{no} GW is sent
forth, because, if we employ Riemann-Fermi coordinates
$y^{j}(\tau)$, any solution of eqs. (\ref{eq:three}) can be
written in the form

\begin{equation} \label{eq:four}
y^{j}(\tau) = a^{j}\tau + b^{j} \quad,
\end{equation}

where $a^{j}$ and $b^{j}$ are constants, and we see that \emph{no}
damping terms due to gravitational radiation are present.

\par From a conceptual point of view, our conclusion is intuitive:
in GR the Newtonian \emph{force} has been substituted by spacetime
\emph{geometry}, so that the bodies ``move freely'' in a
fourdimensional manifold that is \emph{their own} creation
\cite{2}.

\par If there were also \emph{non}-gravitational interactions
(\emph{e.g.}, electromagnetic interactions), the conclusion of the
non-existence of GW's is still valid. A simple, qualitative proof
rests on the fact that the kinematic elements (speeds,
accelerations, time derivatives of accelerations, \emph{etc.}) of
the non-geodesic motions are not different from the kinematic
elements of suitable purely gravitational (geodesic) motions. (See
further $\alpha$), $\beta$), $\gamma$) of \cite{1}).

\vskip1.20cm \noindent \textbf{2bis.} -- Three remarks.  \emph{i})
We have considered the case of a \emph{discrete} system of bodies.
However, our procedure is clearly valid also for a
\emph{continuous} medium. \emph{ii}) The previous formalism
constitutes simply the extension to a system of gravitating masses
of  the well-known formalism concerning the geodesic motions of
\emph{test}-particles in a ``rigidly'' assigned potential field
$g_{jk}$. \emph{iii}) Obviously, formulae \cite{1} $\div$ \cite{4}
hold also for a system of non-gravitating bodies which move freely
in a pseudo-Euclidean spacetime referred to \emph{general}
coordinates.

\vskip1.20cm \noindent \textbf{3.} -- The previous considerations
regard the \emph{exact} formulation of GR. However, it could seem
that in the \emph{linearized} approximation of the theory things
stand otherwise, and that any variation of a material distribution
generates GW's, which are propagated with the velocity of light
\emph{in vacuo}. Now, this approximation -- owing to its formal
resemblance to Maxwell theory -- has created various
misunderstandings, \emph{in primis} about the physical reality of
the GW's.

\par Strictly speaking, the proofs that in the exact GR the GW's
are non-existing objects are sufficient to exclude the adequacy of
any objection based on the linear version. However, the belief in
the physical value of this version is so widespread that a
specific indication of its weak points is quite advisable.

\par As it is well known, in the linearized approximation of GR
one sets:

\begin{equation} \label{eq:five}
g_{jk} \approx \eta_{jk} + h_{jk} \quad,
\end{equation}

where $\eta_{jk}$ is the Minkowskian tensor (in the usual diagonal
form: $1$, $-1$, $-1$, $-1$), and the $h_{jk}$'s are ``small''
deviations from the $\eta_{jk}$'s. The metric tensor
(\ref{eq:five}) has a covariant character \emph{only} under
Lorentz transformations of the coordinates -- and under general,
but ``small'', coordinate transformations: $x^{j} \rightarrow
x^{j} + \xi^{j}(x)$; we have:

\begin{equation} \label{eq:fivex} \tag{5$'$}
h'_{jk} = h_{jk} + \frac{\partial \xi_{j}}{\partial x^{k}} +
\frac{\partial \xi_{k}}{\partial x^{j}} \quad,
\end{equation}

a formula which can also be interpreted as a \emph{gauge
transformation}. (Clearly, the $h_{jk}$'s can be locally
transformed into zero by a \emph{finite} transformation of general
coordinates). Remember that all operations of raising and lowering
of indices are performed with Minkowski tensor $\eta_{jk}$.

\par In a fundamental memoir of 1944 \cite{3}, Weyl gave a new,
original \emph{deduction} of the linear version of GR. It was
obtained with a fresh start, \emph{i.e.} by studying and resolving
the problem of the theoretical existence of a \emph{linear} theory
of gravitation. As main result, \emph{Weyl arrived at
gravitational field equations that are} \textbf{\emph{identical}}
\emph{with the linearized Einsteinian field equations}.

\par If we choose the $\xi_{j}$'s in eqs. (\ref{eq:fivex}) so
that $\gamma_{jk}:= h'_{jk} -(1/2) h'^{n}_{n} \, \eta_{jk}$
satisfies the equations

\begin{equation} \label{eq:fivexx} \tag{5$''$}
\frac{\partial \, \gamma_{j}^{k}}{\partial x^{k}} = 0 \quad,
\end{equation}

the linearized field equations become (as it is known):

\begin{equation} \label{eq:fivexxx} \tag{5$'''$}
\eta^{mn}  \frac{\partial^{2} \gamma_{jk}}{\partial x^{m} \partial
x^{n}} = -2 \kappa \, T_{jk}\quad.
\end{equation}

\par An immediate \emph{consequence} of (\ref{eq:fivexx}) -
(\ref{eq:fivexxx}) is the differential conservation law of
material energy-momentum, which yields also the equations of
motion of bodies:

\begin{equation} \label{eq:six}
\frac{\partial T_{j}^{k}}{\partial x^{k}} = 0 \quad;
\end{equation}

Let us observe explicitly that we have here an \emph{ordinary}
(not a covariant) divergence.

\par Weyl makes a remark that was never previously made: for a ``cloud of
dust'' we have, with obvious notations, $T^{jk}=\mu u^{j}u^{k}$;
accordingly, eqs. (\ref{eq:six}) give the following law of motion
for a ``dust'' particle:

\begin{equation} \label{eq:seven}
\frac{\textrm{d} u^{j}}{\textrm{d} \tau} = 0 \quad,
\end{equation}

and we can say: in the linearized approximation of GR the
gravitational field does \emph{not} exert any force on bodies,
\emph{i.e.} is a ``powerless shadow''.  (Note that the ``cloud of
dust'' is an emblematic instance in GR). This conclusion is
perfectly confirmed by the \emph{exact} theory, for which the
gravitational force on bodies appears ``\emph{only when one
continues the approximation beyond the linear stage}.'' However,
the linearized approximation gives the Poisson-Laplace equation,
and therefore the Newtonian potential $1/r$. (Of course, it is a
partial theory of gravity).

\par \emph{Eqs.} (\ref{eq:seven}) \emph{are mentioned only in
Weyl} \cite{3}. They have a devastating effect on the current
belief in the existence of GW's: indeed, if the motions of the
gravitationally-interacting ``dust'' particles satisfy eqs.
(\ref{eq:seven}), it is indisputable that \emph{no} GW can be
generated by them -- and the customary deduction of the GW's based
on the solution of the linearized homogeneous equations
($T_{jk}=0$) loses any physical meaning. Eqs. (\ref{eq:seven}), as
a consequence of eqs. (\ref{eq:six}), could have been written
immediately after the discovery by Einstein and Grommer in 1927
that the equations of motion of bodies \emph{follow} from Einstein
field equations. We can also say that eqs. (\ref{eq:seven}) are a
simple consequence of the fact that -- as Weyl pointed out in
sect. \textbf{32} of \emph{Raum}-\emph{Zeit}-\emph{Materie}
\cite{4} -- ``$\ldots$ wir befinden uns augenblicklich auf dem
Boden der speziellen Relativi\-t\"atstheorie $\ldots$'';
\emph{i.e.}: the $\textrm{d}s^{2}$ of the linearized approximation
coincides with the Minkowskian $\textrm{d}s^{2}$. An assertion
which could appear amazing to many physicists, who are misled by
the typical instance of a test-particle $\textbf{T}$ (or of a
light-ray \textbf{L}) in a \emph{given}, ``rigid'' field
$\eta_{jk}+ h_{jk}$; it is clear that the motions of $\textbf{T}$
(or of \textbf{L}) are  governed by the customary geodesic
equations in which the $h_{jk}$ play a decisive role. (Remark,
however, that also these kinds of motion do not generate GW's).

\vskip1.20cm \noindent \textbf{3bis.} -- If $\Phi_{k}$ is the
vector potential of the e.m. field $f_{jk}$, the sum

\begin{equation} \label{eq:eight}
\frac{\partial \Phi_{k}}{\partial x^{j}} + \frac{\partial
\Phi_{j}}{\partial x^{k}}
\end{equation}

can \emph{locally }be transformed into zero by means of a gauge
transformation of $\Phi_{k}$. In the gravitational case \emph{all
derivatives} $\partial h_{jk} / \partial x^{m}$ \emph{can  locally
be transformed into zero by virtue of relations} (\ref{eq:fivex}).

\par Whereas in Maxwell theory we have an energy-momentum tensor
of the e.m. field which depends quadratically on the components
$f_{jk}$'s, \emph{no} tensor (different from zero) depending
quadratically on derivatives $\partial h_{jk} / \partial x^{m}$
exists \emph{if} the \emph{gauge relation} (\ref{eq:fivex}) of the
linear version $(L)$ of GR is required. As it was emphasized by
Weyl \cite{3}, with a consideration that is just the counterpart
in $(L)$ of a remark by Levi-Civita (1917) regarding the exact
theory.

\vskip1.20cm \noindent \textbf{3ter.} -- Let us recall that if we
express the equations $R_{jk}- (1/2) g_{jk} R = -\kappa \, T_{jk}$
in a system of \emph{harmonic} coordinates, that are characterized
by the relations $\partial (\sqrt{-g} \, g^{jk}) / \partial
x^{k}=0$, and put $g_{jk} \approx \eta_{jk}+\gamma_{jk}$, eqs.
(\ref{eq:fivexx}) and (\ref{eq:fivexxx}) are obtained immediately.

\par The precise status of the linearized approximation with
respect to the exact theory can be made evident by the following
comparison: \emph{i}) Exact GR: the \emph{covariant} divergences
of both sides of Einstein equations are equal to zero; the
potential field $g_{jk}$ \emph{is} the spacetime -- \emph{ii})
Fundamental property of the linearized approximation: the
\emph{ordinary} divergences of both sides of its field equations
are equal to zero; the approximate version describes a potential
field of a Minkowskian spacetime, referred to a Lorentzian system
of coordinates.

\par This comparison is useful for rendering intuitive the
solution of the problem of the \emph{non}-geodesic motions (see
further \emph{$\alpha$}), \emph{$\beta$}), \emph{$\gamma$}) of
\cite{1}). Assume that the ``dust'' is electrically charged, with
a charge density $\varrho$. In the linearized theory we have the
equations of motion

\begin{equation} \label{eq:sevenx} \tag{7'}
\mu \, \frac{\textrm{d} u^{j}}{\textrm{d} \tau} = \varrho \,
f^{jk} u_{k} \quad,
\end{equation}

\emph{i.e.}, motions that do not generate GW's. In the exact
theory the particles of an electrically neutral ``dust'' obey
geodesic equations that are the strict analogue of eqs.
(\ref{eq:seven}); for a charged ``dust'' the motions of the
particles are non-geodesic, and correspond strictly with the
motions described by the above eqs. (\ref{eq:sevenx}) -- and it is
intuitive that no GW is emitted.

\vskip1.20cm \noindent \textbf{4.} -- As it follows from eqs.
(\ref{eq:fivexxx}), the linearized homogeneous ($T_{jk}=0$)
equations are

\begin{equation} \label{eq:nine}
\eta^{mn} \frac{\partial ^{2} \gamma_{jk}}{\partial x^{m}\partial
x^{n}} = 0 \quad,
\end{equation}

\emph{i.e.} the customary homogeneous d'Alembert equations in
Lorentzian coordinates; the ``waves'' given by \cite{9} represent
undulatory fields in a Minkow\-skian spacetime. However, the
curvature tensor $R_{jklm}$ of the so-called ``TT-waves'' (see in
the sequel) is different from zero -- and this fact seems to
confer them a particular reality. Now, \emph{this} curvature
tensor has an invariant meaning only under \emph{Lorentz}
transformations, and under \emph{infinitesimal} transformations of
general coordinates.  The curvature tensor of a Minkowskian entity
-- as a $\gamma_{jk}$-wave -- is a hybrid notion that mixes
linearized and exact formulations.

\vskip1.20cm \noindent \textbf{5.} -- Let us consider now the
``proof'' of the famous \emph{quadrupole formula}, \emph{e.g.} in
the detailed treatment given by Landau and Lifshitz in sects.
\textbf{101} and \textbf{104} of their book \cite{5}. The central
point of their argumentation is the use of their variant of pseudo
energy-momentum tensor $t^{jk}$ of the gravitational field. In the
exact GR this mathematical object is a false (pseudo) tensor,
because it has a covariant character only under \emph{linear}
transformations (in particular, Lorentz transformations); it can
be transformed into zero at any spacetime point with a suitable
transformation of general coordinates. An analogous conclusion
holds in the linearized approximation, by virtue of eqs.
(\ref{eq:fivex}). In spite of this fact, Landau and Lifshitz wrote
(sect. \textbf{101} of \cite{5}): ``Poss\'edant une energie
determin\'ee, l'onde gravitationnelle cr\'ee elle-m\^eme autour
d'elle un certain champ de gravitation.'' A striking example of a
senseless statement, because only a \emph{true} energy-momentum
tensor can create a gravitational field. (A brilliant proof of the
inadequacy of the very notion of pseudo energy-momentum tensor of
a gravitational field was given by H. Bauer $[$\emph{Phys. Z.},
\textbf{19} (1918) 163$]$, who showed -- with reference to the
first proposed variant of $t^{jk}$ -- that is possible to
introduce spatial coordinates in a \emph{pseudo-Euclidean} world
for which the $t^{jk}$'s are different from zero, and the total
``energy'' is infinite. As an example, Bauer considers the
following coordinates: $\xi^{1}=(1/3)r^{3}$; $\xi^{2}=-\cos
\vartheta$; $\xi^{3}=\varphi$, where $r, \vartheta, \varphi$ are
the usual polar coordinates).

\par For a \emph{plane} undulation which is propagated in the
direction $x^{1}\equiv x$, we can choose in eqs. (\ref{eq:nine})
the coordinates in such a way that $\gamma_{jk}$ is only function
of $x$ and $x^{0}=ct$. Accordingly, eqs. (\ref{eq:nine}) give:

\begin{equation} \label{eq:ten}
 \left(\frac{\partial ^{2}}{\partial x^{2}} - \frac{1}{c^{2}} \frac{\partial ^{2}}{\partial t^{2}}
\right) \gamma_{jk} = 0 \quad;
\end{equation}

it can now be proved that our plane wave is characterized by two
components only, \emph{i.e.} $\gamma_{23}$ and
$\gamma_{22}=-\gamma_{33}$, which are functions of $(x-ct)$: they
are the transverse-transverse (TT) components. The other
components: longitudinal-longitudinal (LL) components and
longitudinal-transverse (LT) components can be eliminated with an
infinitesimal coordinate transformation.

\par Then, the authors compute the pseudo energy-momentum tensor
$t^{jk}$ of the above plane wave, and find that the only component
different from zero is $t^{01}$ (a dot denotes a time derivative):

\begin{equation} \label{eq:eleven}
t^{01} = \frac{c^{2}}{16\,\pi \,G} \, \left[
(\dot{\gamma}_{23})^{2} +
\frac{1}{4}(\dot{\gamma}_{22}-\dot{\gamma}_{33})^{2}\right] \quad.
\end{equation}

By virtue of eqs. (\ref{eq:fivex}), all these derivatives can be
locally transformed into zero.

\par In sect. \textbf{104} of \cite{5} Landau and Lifshitz
investigate the \emph{weak} gravitational field generated by
bodies in \emph{slow} motions. They write (cf. our eqs.
(\ref{eq:fivexxx}), (\ref{eq:fivexx}) and (\ref{eq:six})):

\begin{equation} \label{eq:twelve}
\frac{1}{2} \eta^{mn} \frac{\partial ^{2}\gamma_{jk}}{\partial
x^{m}\partial x^{n}} = - \frac{8\,\pi \,G}{c^{4}} (\mu \, u_{j} \,
u_{k}) \quad,
\end{equation}

\begin{equation} \label{eq:thirteen}
\frac{\partial \gamma^{jk}}{\partial x^{k}} = 0 = \frac{\partial
}{\partial x^{k}} (\mu \, u^{j} \, u^{k}) \quad.
\end{equation}

The authors are interested in the following solution of eqs.
(\ref{eq:twelve}) -- the meaning of symbols is obvious:

\begin{equation} \label{eq:fourteen}
\gamma^{jk} = \frac{4G}{c^{4}} \int [\mu \, u^{j} \,
u^{k}]_{t-\frac{\textrm{R}}{c}} \, \frac{\textrm{d}V}{\textrm{R}}
\quad;
\end{equation}

by taking into account the smallness of the speeds of the bodies
of our physical system, eqs. (\ref{eq:fourteen}) can be
approximated, for the very far field, as follows:

\begin{equation} \label{eq:fifteen}
 \gamma^{jk} = \frac{4G}{c^{4}} \,
  \frac{1}{\textrm{R}_{0}}
\int
 [\mu \, u^{j} \, u^{k}]_{t-\frac{\textrm{R}_{0}}{c}} \,
\textrm{d}V \quad,
\end{equation}

if $\textrm{R}_{0}$ is the distance from the coordinate origin,
situated in a point within the spatial region of the system. Using
the second set of eqs. (\ref{eq:thirteen}), one can compute
various relations among the integrals of (\ref{eq:fifteen}); one
finds $(\alpha, \beta = 1, 2, 3)$:

\begin{equation} \label{eq:sixteen}
\gamma_{\alpha \beta} = \frac{2G}{c^{4}\textrm{R}_{0}} \,
\frac{\partial ^{2}}{\partial t^{2}} \int
\mu\left(\textrm{\textbf{x}}, t-\frac{\textrm{R}_{0}}{c}\right) \,
x_{\alpha}\, x_{\beta} \, \textrm{d}V \quad.
\end{equation}

\par At \emph{large} distances from the physical system, and
inside \emph{small} regions, the GW is practically a \emph{plane}
wave. Accordingly, we can compute the energy flow emitted by the
system in $x_{1}$-direction by means of eq. (\ref{eq:eleven});
from eqs. (\ref{eq:sixteen}) we have:

\begin{equation} \label{eq:seventeen}
\gamma_{23} = \frac{2G}{3c^{4}\textrm{R}_{0}} \, \ddot{D}_{23}
\quad; \quad \quad \gamma_{22}-\gamma_{23} =
\frac{2G}{c^{4}\textrm{R}_{0}} \, (\ddot{D}_{22} -\ddot{D}_{33})
\quad,
\end{equation}

where

\begin{equation} \label{eq:eighteen}
D_{\alpha \beta}:= \int \mu\left(\textrm{\textbf{x}},
t-\frac{\textrm{R}_{0}}{c}\right) \left[ 3 \, x_{\alpha}\,
x_{\beta} - \delta_{\alpha \beta}\, x_{\gamma} \, x_{\gamma}
\right] \, \textrm{d}V
\end{equation}

is the quadrupole moment of masses. Then, the pseudo tensor
$t^{01}$ of eq. (\ref{eq:eleven}) can be written:

\begin{equation} \label{eq:nineteen}
c\,t^{01} = \frac{G}{36 \, \pi \, c^{5}\textrm{R}_{0}^{2}} \left[
\left(\frac{\dddot{D}_{22}-\dddot{D}_{33}}{2} \right)^{2} -
\left(\dddot{D}_{23} \right)^{2}\right] \quad.
\end{equation}

One derives easily from (\ref{eq:nineteen}) the radiation emitted
in an arbitrary direction, and then the total radiation emitted in
all directions, \emph{i.e.} the energy lost by our system in a
second:

\begin{equation} \label{eq:twenty}
- \frac{\textrm{d}\mathcal{E}}{\textrm{d}t} = \frac{G}{45 \,
c^{5}} \, \dddot{D}_{\alpha \beta} \, \dddot{D}_{\alpha \beta}
\quad.
\end{equation}

I have reported almost literally, with inessential modifications,
some passages of the treatment by Landau and Lifshitz \cite{5}.

\par The authors give now the result of a computation which
anticipates the computations concerning the celebrated binary B
PSR1913+16. They consider two bodies which interact according to
Newton law and describe circular orbits. If $m_{1}$, $m_{2}$ are
their masses, $r$ their distance, and $T=2\,\pi/\omega$ their
revolution period, we obtain:

\begin{equation} \label{eq:twentyone}
- \frac{\textrm{d}\mathcal{E}}{\textrm{d}t} = \frac{32\,G}{5 \,
c^{5}} \, \left(\frac{m_{1}m_{2}}{m_{1}+m_{2}} \right)^{2} \,
r^{4} \, \omega^{6} \quad;
\end{equation}

at this end, it is necessary to perform in eqs.
(\ref{eq:eighteen}) the passage from the continuous
$\mu\left[\textrm{\textbf{x}}, t-(\textrm{R}_{0}/c\right]$ to its
corresponding discrete expression by means of Dirac's
delta-distributions (masses $m_{1}$, $m_{2}$ are considered as
\emph{pointlike}). Since, clearly:

\begin{equation} \label{eq:twentytwo}
\omega^{2}r^{3} = G(m_{1}+m_{2}) \quad; \quad \mathcal{E}=-
\frac{G\,m_{1}m_{2}}{2r} \quad,
\end{equation}

we have:

\begin{equation} \label{eq:twentythree}
\frac{\textrm{d}r}{\textrm{d}t} = \frac{2r^{2}}{G\,m_{1}m_{2}} \,
\frac{\textrm{d}\mathcal{E}}{\textrm{d}t} = -
\frac{64G^{3}m_{1}m_{2}\, (m_{1}+m_{2})}{5c^{5}r^{3}} \quad,
\end{equation}

which gives the approaching speed of the two bodies as it follows
from the energy loss due to gravitational radiation.

\vskip1.20cm \noindent \textbf{6.} -- The results of  previous
sect.\textbf{5} are \emph{physically} meaningless. Indeed, since
we know that the equations

\begin{equation} \label{eq:twentyfour}
\frac{\partial}{\partial x^{k}} \, (\mu \, u^{j} u^{k}) = 0\quad;
\end{equation}

have as a consequence $\textrm{d}u^{j}/ \textrm{d}\tau =0$ (sect.
\textbf{3}), it is \emph{not} allowed to postulate in the
\emph{linearized} approximation of GR the existence of the action
on bodies of Newton force; actually, the gravitational force on
them appears only \emph{beyond} the linear stage. Further, eqs.
(\ref{eq:twentyfour}) tell us that no gravitational motion
generates GW's, and therefore the undulatory solutions of eqs.
(\ref{eq:nine}) are destitute of a physical reality.

\par In the \emph{exact} formulation of GR the ``dust'' particles
describe \emph{geodesic} lines -- and therefore cannot emit GW's
(cf. sect. \textbf{2}). Further, as it was pointed out by Weyl --
see p.268 of \cite{4} -- it is always possible to choose a
coordinate system for which two gravitating bodies in relative
motion are \emph{both} at rest. This Weylian observation clarifies
very well the conceptual meaning of the computations by Thirring
and Lense: instances of effects of Einsteinian ``dragging''
forces, not  very different in nature from the Newtonian
centrifugal and Coriolis forces. In general relativity \emph{no}
kinematic parameter (velocity, acceleration, time derivative of
acceleration, \emph{etc}.) has an invariant character. On the
contrary, in Maxwell theory only the reference frames in
rectilinear and uniform relative motions are physically
equivalent.

\vskip1.20cm \noindent \textbf{7.} -- If we apply the treatment of
sect. \textbf{2} to the linear version of GR, where
$\textrm{d}s^{2}$ coincides with Minkowskian $\textrm{d}s^{2}$, we
find immediately, in lieu of eqs. (\ref{eq:three}), the simple
equations

\begin{equation} \label{eq:twentyfive}
\frac{\textrm{d}^{2}q^{j}}{\textrm{d}\tau^{2}} =0 \quad,
\end{equation}

\emph{i.e.}, eqs. (\ref{eq:seven}).

\vskip1.20cm \noindent \textbf{8.} -- Sects. \textbf{3} to
\textbf{7} (with the exception of the formulae by Landau and
Lifshitz \cite{5}) will result unpalatable to many readers, owing
to a diffuse opinion that the Weylian linear version of GR
\cite{3} does not coincide exactly with the current linearized
approximation. Of course, this is not true, as a simple inspection
of the corresponding formulae can easily show. (The only,
inessential, difference is that in \cite{3} the ``smallness'' of
the $h_{jk}$'s is not postulated).

\par  I emphasize again that there are two \emph{different} applications of the
 approximate equality $g_{jk} \approx \eta_{jk} + h_{jk}$. The
 application No.1 concerns the computations of the geodesic
 motions of test-particles and light-rays in ``external''
 manifolds that are weakly pseudo-Riemannian. The application No.2
 regards the approximation of the Einsteinian field equations; it
 gives eqs. (\ref{eq:fivexx}) and (\ref{eq:fivexxx}): \emph{i.e.},
 the linear version of GR (Minkowskian spacetime).

 \par Also the results of sects. \textbf{2} and \textbf{2bis} can
 give rise to some perplexity, because they are at variance with
 current ideas on GW's. As a matter of fact, GR does not admit the
 \emph{physical} existence of GW's -- as it was first proved by
 Levi-Civita in 1917. \emph{And} \textbf{\emph{experience}}
 \emph{continues to confirm the validity of this theorem}
 \cite{6}. A very simple and qualitative proof of it runs as
 follows (for an analytical support see \cite{7}). Consider the
 general concept of \emph{wave} (without any adjective). There
 exist waves that are undulatory perturbations of material media
 (as air, water, \emph{etc}.; formerly, cosmic ether), and waves
 that are undulatory perturbations of fields \emph{in vacuo} with
 respect to an infinite class of physically privileged reference
 frames, as the Lorentzian frames of special relativity. More
 generally, we can have field waves \emph{in vacuo} with
 respect to an infinite class of physically privileged coordinate
 systems of a ``rigidly given'' pseudo-Riemannian manifold,
 endowed with \emph{uniformity} properties. Now, in the
 \emph{exact} formulation of GR \emph{the metric tensor} $g_{jk}$ \textbf{\emph{is}}
 \emph{the spacetime}; thus, an undulatory $g_{jk}$ \emph{lacks}
 of any spatio-temporal substrate (as Minkowski spacetime or
 uniform pseudo-Riemannian manifold) through which it can be
 propagated. Consequently, it is doomed to be a property of
 \emph{some} systems of general coordinates; \emph{a change of
 coordinates can impair its undulatory character, and give it
  an arbitrary velocity of propagation}. In short: it is only a
  mathematical object with a zero energy and without physical
  reality -- even if its curvature tensor is different from zero.

\vskip1.20cm \noindent \textbf{9.} -- Let us consider in Minkowski
spacetime of special relativity a tensor field $\varphi_{jk}(x) \,
[=\varphi_{kj}(x)]$, $(j,k=0, 1, 2, 3)$, satisfying the following
equations:

\begin{equation} \label{eq:twentysix}
\eta^{mn} \frac{\partial ^{2}\varphi_{jk}}{\partial x^{m} \partial
x^{n}} = S_{jk} \quad,
\end{equation}

\begin{equation} \label{eq:twentyseven}
\frac{\partial \varphi_{jk}}{\partial x_{k}} = 0 \quad, \quad
\left( \Rightarrow \frac{\partial S_{jk}}{\partial x_{k}}  = 0
\right) \quad;
\end{equation}

$\varphi_{jk}$ and its source $S_{jk}$ are of an indefinite
physical nature. Eqs. (\ref{eq:twentysix}) and
(\ref{eq:twentyseven}) are invariant under the \emph{gauge
transformations}

\begin{equation} \label{eq:twentyeight}
\varphi_{jk} \rightarrow \varphi'_{jk} = \varphi_{jk} +
 \frac{\partial \xi_{j}}{\partial x^{k}} + \frac{\partial \xi_{k}}{\partial
 x^{j}} - \eta_{jk} \, \frac{\partial \xi_{m}}{\partial x_{m}}
 \quad,
\end{equation}

where the four functions $\xi_{j}(x)$ satisfy the homogeneous
d'Alembert equations

\begin{equation} \label{eq:twentynine}
\frac{\partial ^{2}\xi_{j}}{\partial x_{k} \partial x^{k}} = 0
\quad;
\end{equation}

(Of course, $S_{jk}$ is a gauge invariant quantity, as the current
$s_{k}$ of Maxwell theory).

\par If we put $\varphi_{jk} = \gamma_{jk}$, and $S_{jk}= -2
\kappa\, T_{jk}$, we re-obtain the equations of the linearized
approximation of GR. We see in the clearest way that \emph{in this
approximation we are dealing with the} \textbf{\emph{special}}
\emph{relativity} \cite{4}.

\par Some authors affirm that the nonlinearity of the Einsteinian
field equations is the only responsible for the fact that these
equations have as a consequence the equations of motions of
bodies. A false opinion: indeed, such a consequence exists also in
the linearized approximation of the theory.

\vskip1.60cm
\begin{center}
\noindent \small \emph{\textbf{Historical appendix}}
\end{center}
\normalsize \noindent \vskip0.60cm

\par In the current literature -- even in a literature of
historical character (see, \emph{e.g.}, \cite{8} -- Weyl's memoir
\cite{3} is strangely ignored. An exiguous minority of
theoreticians have read it, and have concluded erroneously that
Weyl's linear theory contradicts the usual linearized
approximation of GR. Maybe, Weyl's observation that in the linear
version of GR the gravitational field is a powerless shadow
(because it does not exert any force on bodies) can have misled
many cursory readers. In reality -- as I have pointed out in sect.
\textbf{3} --, Weyl's result could have been discovered
immediately after 1927, when the geodesic \emph{principle} became
the geodesic \emph{theorem}.

\par Another strange fact is the current overlooking of Hilbertian
repulsive effect \cite{9}, with its important consequences on the
behaviour of geodesic lines of test-particles and light-rays in a
given Schwarzschild manifold. An extraneous factor could have
played a role in this neglect: the diffuse belief in surprising
properties of a well-known ``soft'' geometrical singularity.

\par It is interesting that \emph{Hilbert} \cite{9} \emph{does not
mention the GW's}, and utilizes the linearized approximation of GR
only for a perturbative proof of Serini's theorem \cite{10}: the
non-existence of \emph{regular} time-independent solutions of
$R_{jk}=0$, that become pseudo-Euclidean at spatial infinity;
\emph{i.e.}, the unique regular solution of this kind is the
Minkowskian one.

\par Finally, I wish to recall a significant remark by Hilbert \cite{9}
on the physical meaning of any statement (\emph{Aussage}) in
general relativity. He emphasized that in GR a given statement has
a real physical meaning \emph{only if} it has an invariant
character under any whatever transformation of general
coordinates. An analogous criterion holds obviously for the
properties (\emph{Eigenschaften}). (Thus, \emph{e.g.}, the wave
character of a $g_{jk}$, and its velocity of propagation, are
unphysical properties, see sect. \textbf{8}).

\par A geometric comparison: in the differential geometry of
curves and surfaces a given statement, or a given property, have a
real geometric meaning \emph{only if} they are independent of the
choice of the coordinates.

\par Hilbert wrote \cite{9}: ``Dem Wesen des neuen Relativit\"atsprinzipes
$[$\emph{i.e.}, of GR$]$ entsprechend m\"ussen wir $[\ldots]$ die
Invarianz nicht nur f\"ur die allgemeinen Gesetze der Physik
verlangen, sondern auch jeder Einzelaussage $[$to any single
statement$]$ in der Physik den invarianten Charakter zusprechen,
falls sie einen physikalischen Sinn haben soll -- im Einklang
damit, da\ss{ }jede physikalische Tatsache letzen Endes durch
Lichtuhren, d.h. durch Instrumente von \emph{invariantem}
Charakter feststellen sein mu\ss. Genau so wie in der Kurven - und
Fl\"achentheorie eine Aussage, f\"ur die die Parameterdarstellung
der Kurve oder Fl\"ache gew\"ahlt ist, f\"ur die Kurve oder
Fl\"ache selbst keinen geometrischen Sinn hat, wenn nicht die
Aussage gegen\"uber einer beliebigen Transformation der Parameter
invariant bleibt oder sich in eine invariante Form bringen l\"a\ss
t, so m\"ussen wir auch in der Physik eine Aussage, die nicht
gegen\"uber jeder beliebigen Transformation des Koordinatensystems
invariant bleibt, als \emph{physikalisch sinnlos} bezeichnen.''

%% \newpage

\end{document}